\documentclass{article}

\usepackage[numbers]{natbib}

\usepackage[final]{neurips_2024}

\usepackage[utf8]{inputenc} 
\usepackage[T1]{fontenc}    
\usepackage{hyperref}       
\usepackage{url}            
\usepackage{booktabs}       
\usepackage{amsfonts}       
\usepackage{nicefrac}       
\usepackage{microtype}      
\usepackage{colortbl}
\usepackage{xcolor}
\usepackage{amsmath}
\usepackage{amssymb}
\usepackage{ amssymb }
\usepackage{xcolor}
\usepackage{booktabs}
\usepackage{makecell} 

\newcommand{\bfc}{\mathbf{c}}

\newcommand{\bfn}{\mathbf{n}}

\newcommand{\bft}{\mathbf{t}}
\newcommand{\bfu}{\mathbf{u}}

\newcommand{\bfx}{\mathbf{x}}

\usepackage{subfig}
\usepackage{graphicx}
\graphicspath{{./Figures/}}

\title{Developing a Foundation Model for Predicting Material Failure}

\author{
  Agnese Marcato\And
  Javier E. Santos \And
  Aleksandra Pachalieva \And  
  Kai Gao \And 
  Ryley Hill \And 
  Esteban Rougier \And 
  Qinjun Kang \And 
  Jeffrey Hyman \And 
  Abigail Hunter \And 
  Janel Chua \And 
  Earl Lawrence \And 
  Hari Viswanathan \And 
  Daniel O'Malley
}

\begin{document}

\maketitle
\begin{center}
\vspace{-.75cm}
    Los Alamos National Laboratory\\
    Los Alamos, NM, US
    \vspace{.75cm}
\end{center}

\begin{abstract}

Understanding material failure is critical for designing stronger and lighter structures by identifying weaknesses that could be mitigated, predicting the integrity of engineered systems under stress to prevent unexpected breakdowns, and evaluating fractured subsurface reservoirs to ensure the long-term stability of the reservoir walls, fluid containment, and surrounding geological formations. Existing full-physics numerical simulation techniques involve trade-offs between speed, accuracy, and the ability to handle complex features like varying boundary conditions, grid types, resolution, and physical models. While each of these aspects is important, relying on a single method is often insufficient, and performing a comprehensive suite of simulations to capture variability and uncertainty is impractical due to computational constraints.
We present the first foundation model specifically designed for predicting material failure, leveraging large-scale datasets and a high parameter count (up to 3B) to significantly improve the accuracy of failure predictions. In addition, a large language model provides rich context embeddings, enabling our model to make predictions across a diverse range of conditions. Unlike traditional machine learning models, which are often tailored to specific systems or limited to narrow simulation conditions, our foundation model is designed to generalize across different materials and simulators. This flexibility enables the model to handle a range of material properties and conditions, providing accurate predictions without the need for retraining or adjustments for each specific case. Our model is capable of accommodating diverse input formats, such as images and varying simulation conditions, and producing a range of outputs, from simulation results to effective properties. It supports both Cartesian and unstructured grids, with design choices that allow for seamless updates and extensions as new data and requirements emerge.
Our results show that increasing the scale of the model leads to significant performance gains (loss scales as $N^{-1.6}$, compared to language models which often scale as $N^{-0.5}$). This model represents a key stepping stone to advancing predictive capabilities of material science and related fields.

\end{abstract}


\section{Introduction}

Fracturing is a common phenomenon observed across a wide range of scientific domains and engineering applications, including subsurface geology~\cite{hyman2016understanding}, earthquake rupture~\cite{husseini1975fracture}, pipeline integrity~\cite{mohtadi2019effects}, steel and concrete structures~\cite{golewski2023phenomenon}, response of human-made systems to impulsive loads~\cite{kalthoff1986fracture}, material design~\cite{launey2009fracture}, high explosives performance~\cite{hobbs2014ignition}, and biological structures such as bones~\cite{gupta2008fracture}, just to name a few.  Accurately simulating how these fractures interact with the surrounding stress field and how they may propagate under strain is crucial for advancements in these fields, but presents significant challenges due to the high computational costs (full physics simulators) and complexity (highly heterogenous materials) involved.

Machine learning, and especially natural language processing, has seen remarkable progress through the development of large-scale models trained on big and diverse datasets~\cite{brown2020language,kenton2019bert,chowdhery2023palm}. Scientific domains could similarly benefit from such approaches. However, scientific data presents unique challenges: it is expensive to generate, challenging for quality control, and often multimodal, encompassing various formats and simulation types. Despite these hurdles, some scientific fields have successfully applied large-scale modeling techniques. In protein folding prediction, models like AlphaFold~\cite{jumper2021highly} have achieved unprecedented accuracy in predicting 3D protein structures. In drug discovery, machine learning models have accelerated the identification of potential therapeutic compounds~\cite{stokes2020deep}. Similarly, in climate modeling, large-scale models have improved the accuracy of long-term climate predictions~\cite{palmer2008toward, bodnar2024aurora}, and in computational chemistry, they have enhanced the understanding of molecular interactions~\cite{rupp2012fast,ahmad2022chemberta}. These successes demonstrate the potential of large-scale models in addressing complex scientific challenges.

In this work, we introduce a multimodal foundation model for predicting material failure. Our  model can perform multiple tasks related to fracture prediction -- namely predicting how long it will take a material to fail after loading and the pattern of fractures that cause the material to fail. The model is trained on data from three different fracture simulators in a curriculum-style approach. The first is a rule-based model that captures first order fracture growth behavior and can generate vast amounts of data on the fly. The second is a quasi-static phase-field fracture simulator that represents the fracture evolution without considering the inertia of the system. The third uses the combined finite-discrete element method to capture the full dynamics of a system under load, with the corresponding fracture initiation, propagation, and arrest, coupled with the resolution of the interaction between the discrete blocks or parts. This numerical model, while offering the highest fidelity, is also the most computationally demanding.

Our framework serves as a guide for future work by showcasing how a single foundation model can progressively learn from datasets with varying levels of physical complexity -- ranging from simple, rule-based simulations to high-fidelity, finite-discrete element models. The model handles diverse input formats, including both Cartesian and unstructured grids, and supports multiple prediction tasks such as fracture patterns and time-to-failure across different materials and loading conditions. This unified approach, which accommodates varied simulation environments and material behaviors, demonstrates the potential of foundation models to address complex physical systems in ways that traditional, system-specific models cannot.

We employ a transformer-based architecture derived from the Senseiver~\cite{santos2023development} model, training models with up to 3 billion parameters. Our hypothesis is that by scaling to billion-parameter models, coupled with our architecture, the model will exhibit foundational properties -- similar to those observed in large language models -- enabling it to predict material failure across varying material types and boundary conditions. While still in early stages, this approach shows promise for generalizing beyond specific datasets and simulation conditions, making it more versatile for a range of scientific applications. We  describe the techniques required to effectively conduct the training of data-driven models at this scale using scientific data. We also study how the model performance scales both in terms of the data size and the number of parameters. We find that model accuracy improves much more rapidly as parameters increase than is found in language models, suggesting that scientific data may have a structure that can be accurately modeled using fewer parameters than language models.

\begin{figure}
    \centering
    \includegraphics[width=\textwidth]{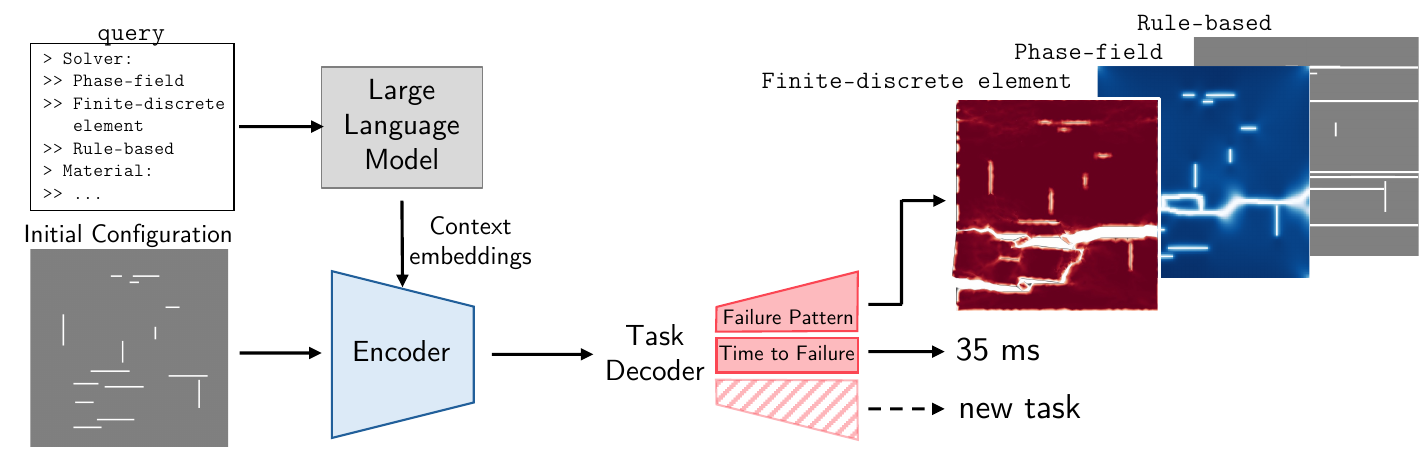}
    \caption{Schematic of our foundation model for predicting material failure.}
    \label{fig:workflow}
\end{figure}

\section{Model Architecture}

Our model is built on an encoder-decoder framework designed to handle diverse input types and perform multiple prediction tasks, Figure~\ref{fig:workflow}. While encoder-decoder architectures are common in other domains, applying them in this context -- where the model must process both structured and unstructured grids and predict complex outputs like fracture patterns and time-to-failure -- has not been widely explored. This combination of handling various grid formats and predicting multiple fracture-related phenomena in a unified framework represents a novel approach in the field of material science, addressing the unique challenges of fracture failure prediction. The main features of the model are summarized below, while detailed technical specifications are provided in Appendix~\ref{sec:architecture}.

\subsection{Encoder: Flexible Input Size}

The encoder leverages cross-attention mechanisms to process all inputs as 1D sequences (with multidimensional position encodings -- see Appendix A.1), making it agnostic to domain size as shown in the example predictions of Figure \ref{fig:diff_sizes}. This layer scales linearly with input size. A key aspect of the encoder is its positional embedding, which maps grid bins to their respective center points. This is crucial as it allows the model to handle both structured (Cartesian) and unstructured grids (e.g., triangular elements in our case).

\subsubsection{Incorporating Additional Context}

Traditionally, context has been integrated using categorical embeddings or trained latent spaces. In contrast, we propose using a large language model (LLM)-based encoder, offering a more versatile and future-proof framework. By utilizing rich, pre-trained representations from Meta’s LLaMA-3 8B, our architecture supports a broader range of inputs, including boundary conditions, material properties, and simulation scenarios. This LLM-based encoder not only retains compatibility with traditional categorical classes but also expands the model’s ability to generalize across diverse input types. This architecture ensures adaptability and longevity, allowing us to continuously refine the base model as new data and tasks emerge. Details about the creation of the LLM embeddings can be found in Appendix~\ref{sec:creation_emb}.

\subsection{Decoder: Multiple tasks}

When predicting material failure, two factors are key: the final fracture pattern and the time it takes to reach that point. Our decoder addresses these aspects by generating the final fracture pattern from an initial condition and predicting the time-to-failure, typically measured in milliseconds. Using trainable task embeddings, the model outputs both an image of the failure pattern, matching the input size, and a floating-point number representing the time until failure under specific conditions. This process is illustrated in Figure \ref{fig:workflow}.

\section{Pre-training}

\subsection{Hardware and Training set-up}

We trained our model in the Venado supercomputer housed at Los Alamos National Laboratory. Named after the Venado Peak (a mountain near Taos, New Mexico), the exascale supercomputer system consists of 2,560 Nvidia Grace Hopper superchips. Each chip integrates a Grace CPU with a Hopper architecture GPU, featuring 44 Arm cores. This hardware is well-suited for large-scale training tasks due to its high computational memory and fast processing capabilities \cite{grace_hopper}. We used 16 nodes, each equipped with four H100 GPUs, totaling 64 H100 GPUs for our training setup.  To efficiently distribute the training across these GPUs, we employed DistributedDataParallel (DDP) with PyTorch Lightning~\cite{Falcon_PyTorch_Lightning_2019}. DDP was selected for its excellent scalability and straightforward implementation. The ample memory of the H100 allowed the entire model to be loaded on each GPU, enabling full utilization of parallel processing capabilities. In this setup, each GPU independently processed different mini-batches with minimal communication overhead, resulting in faster and more efficient training without the need for complex configurations.

\subsection{Pre-training objective}

Pre-training played a crucial role in our model's development. By exposing the model to a large body of uncurated data, it began to focus on important patterns and key structures, which are essential for generalization. During pre-training, we generated data on the fly to ensure that the model learns the tasks effectively. The objective of this first-phase is for the model to understand how to produce the earliest failure pattern of the material and to approximate the physical time it takes to reach that pattern. 
\begin{equation}
\label{eq:loss}
\mathcal{L} = \frac{1}{N} \sum_{i=1}^{N} \left[-y_{f_i} \log(\hat{y}_{f_i}) - (1 - y_{f_i}) \log(1 - \hat{y}_i)\right] +  \frac{1}{N} (\hat{y_t} - y_t)^2
\end{equation}
The loss function \(\mathcal{L}\) consists of two main terms: the first term is the binary cross-entropy loss, averaged over \(N\), the number of elements in the grid, where \(y_{f_i}\) represents the true label indicating the presence or absence of a fracture at the \(i\)th element, and \(\hat{y}_{f_i}\) is the predicted probability of the failed fracture pattern at that element. The second term is the mean squared error between the predicted time to failure \(\hat{y_t}\) and the true time to failure \(y_t\).

\subsection{Rule-based algorithm for the creation of data}
The pre-training of the model is based on the data coming from a rule-based algorithm emulating the fracturing of materials. 
The algorithm grows the fractures in the direction normal to the load. For example if the material is pulled vertically, the fractures grow horizontally following two growth rules: (-) ``X'' growth: where the fractures can break through other fractures and continue growing. (-) ``T'' growth: where the fractures stop their expansion when they collide with another fracture.
These two scenarios, even if implemented in a rule-based algorithm, are realistic for material fracturing~\cite{zhu2023impacts}. For example, these models have been successful in predicting fracture growth in sea ice~\cite{vevatne2014fracture}. The algorithm implementation is detailed in Appendix~\ref{sec:wannabe} together with examples of the generated failure patterns (Figure~\ref{fig:wannabe}). The rule-based algorithm generates new realizations within milliseconds, allowing us to produce data dynamically during training. Instead of relying on a fixed dataset, we create new realizations at each optimization step. These rule-based simulations run on 72 cores the Grace CPU while the model is trained on the Hopper GPU.

\subsection{Scaling study}

We conducted a series of experiments to evaluate how the number of parameters influences the value of the final loss and, consequently, its accuracy. To scale our model, we increased the number of channels in both the encoder and decoder, as well as the number of attention blocks. During training, we varied the position of the fractures within the domain, effectively generating an infinite dataset via on-the-fly data generation.

\begin{figure}
    \centering
    \includegraphics[width=\textwidth]{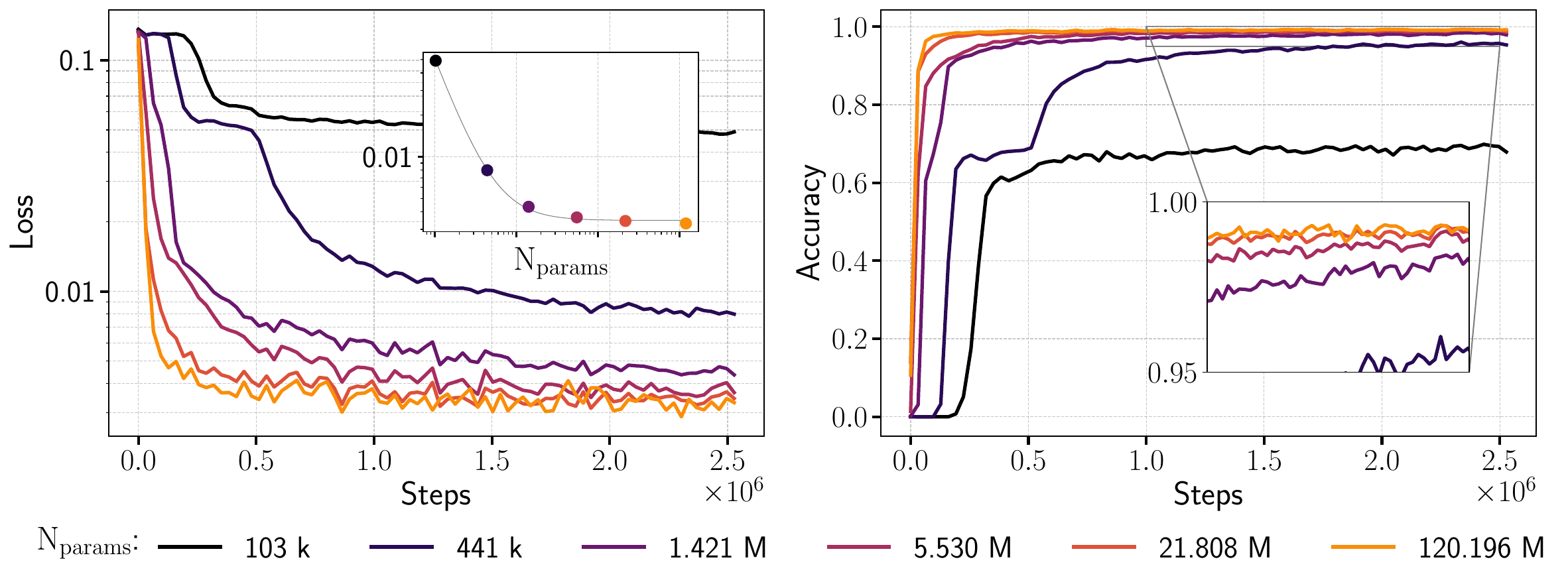}
    \caption{Training loss (Eq. \ref{eq:loss}) and accuracy across models with varying parameter counts. Left: Loss decreases consistently with increased parameters, as shown in the inset plot highlighting the relationship between parameter count ($N_{params}$) and loss. Right: Accuracy improves significantly with larger models, nearing 100\% for models with over 1 million parameters. Insets detail performance differences at high accuracy levels. }
    \label{fig:loss_accuracy}
\end{figure}

\begin{figure}
    \centering
    \includegraphics[width=0.68\textwidth]{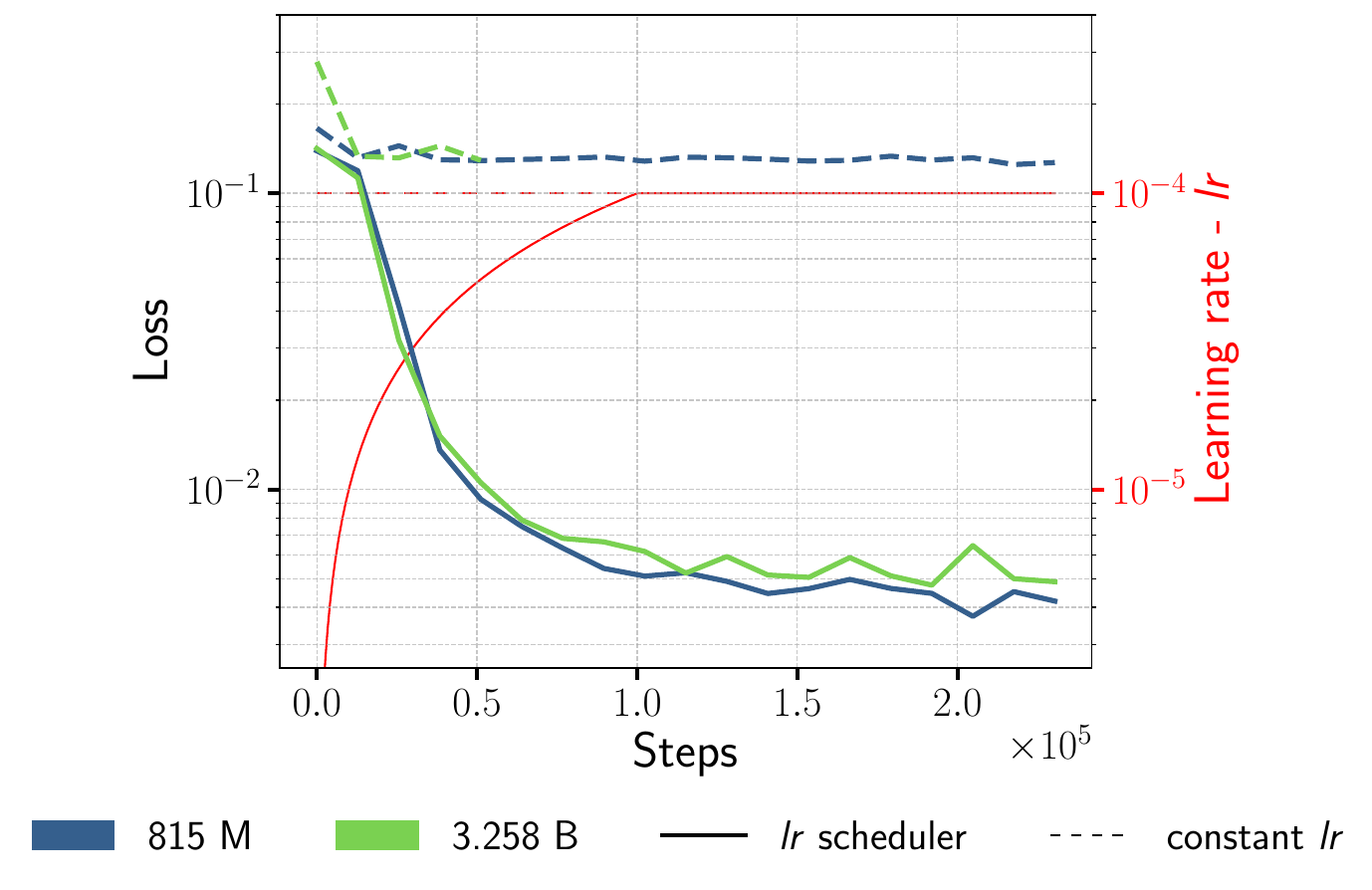}
    \caption{Loss curves illustrating the impact of a warm-up phase on training performance for two large model sizes (815M and 3.258B parameters). Models trained with a warm-up phase (solid lines) show significantly improved convergence compared to those trained with constant learning rates (dashed lines). The results indicate that a warm-up phase is crucial for effectively training larger models with scientific data.}
    \label{fig:scheduler}
\end{figure}

In Figure~\ref{fig:loss_accuracy} we present the loss curves and accuracy during training, parameterized by model size, which ranges from 103k to 120M parameters. Detailed hyperparameters for each of these models are provided in Appendix~\ref{sec:scaleup}. By increasing the model size, the final training loss decreases following the relationship:
\begin{equation}
    \mathcal{L} = \frac{3.91\times 10^6}{\mathrm{{N_{params}}^{1.58}}} + 3.48 \times 10^{-3}.
\end{equation}
This trend is illustrated in the left panel of Figure~\ref{fig:loss_accuracy}. We further scaled our model up to 800 million and 3.2 billion parameters, as shown in Figure~\ref{fig:scheduler}. When trained with a constant learning rate, as done in experiments with smaller models, the loss plateaued at a high value and did not decrease further. To address this, we implemented a learning rate scheduler~\cite{gotmare2018closer}  including a warm-up phase where the learning rate increased from 0 to a set point of $10^{-4}$ over the first 100,000 steps, followed by a decay according to a cosine schedule. With the scheduler in place, the model was able to train successfully.

\subsection{Pre-training with LLM embeddings context}

We conducted a proof-of-concept experiment to evaluate the use of LLM embeddings, leveraging pre-training data generated by the rule-based algorithm. We hypothesize that incorporating LLM embeddings will enhance the model's ability to differentiate between various loading conditions, fracture orientations, and material types. By embedding these contextual elements during training, we aim to enhance the model's ability to generalize across diverse scenarios, improving its predictive accuracy and versatility. This integration provides a more nuanced understanding of the physical parameters involved, potentially leading to more robust predictions across a variety of material failure conditions. In the long term, these language model embeddings will provide a flexible way to describe simulator types, boundary conditions, material properties, etc.

\begin{figure}[htb!]
    \centering
    \begin{minipage}{0.3\textwidth}
        \centering
\begin{tabular}{@{}ccc@{}}
\toprule
        & Horizontal & Vertical \\ 
        T & \cellcolor[HTML]{B0C4DE}   \makecell{96.95\% 
        } & \cellcolor[HTML]{87CEFA}   \makecell{98.54\% 
        } \\
        X & \cellcolor[HTML]{AFEEEE}   \makecell{96.74\% 
        } & \cellcolor[HTML]{FF6666}   \makecell{98.33\% 
        } \\
\bottomrule
\end{tabular}
 \captionof{table}{Accuracy on the failure pattern for 1000 examples, blue shades: training embeddings, red: testing embedding.}
\label{tab:LLM-fp}
    \end{minipage}%
    \hspace{0.05\textwidth}
    \begin{minipage}{0.5\textwidth}
        \centering
\includegraphics[width=\textwidth]{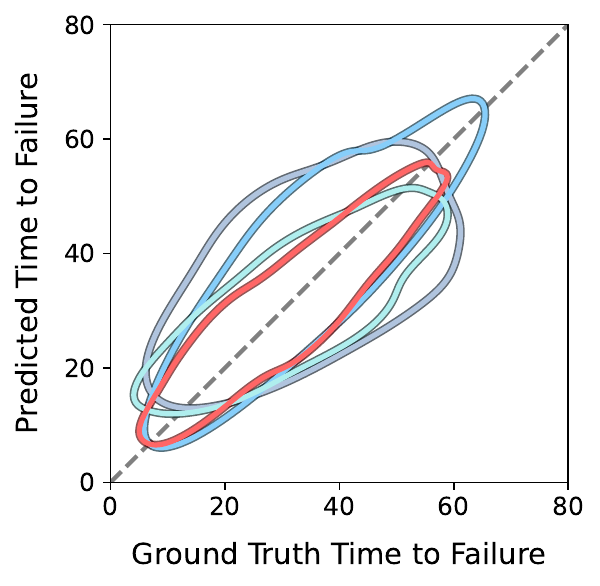}
        \caption{Time to failure parity diagram, the colors refer to Table~\ref{tab:LLM-fp}.}
\label{fig:LLM-ttf}
    \end{minipage}
\end{figure}
\begin{figure}
    \centering
\includegraphics[width=0.75\textwidth]{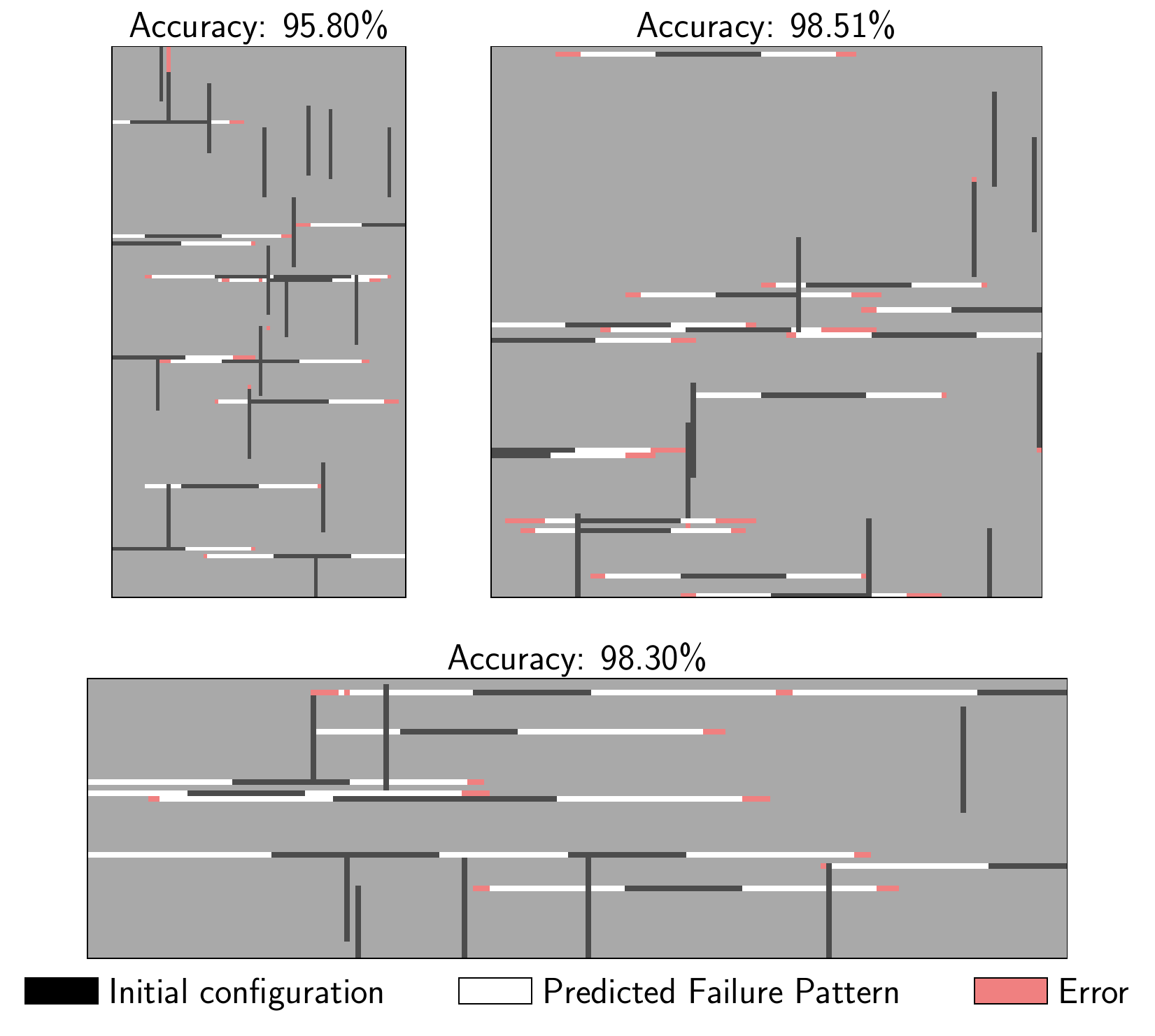}
\caption{Examples of model predictions of different domain sizes and with varying number of initial fractures.}
\label{fig:diff_sizes}
\end{figure}

The rule-based model simulates two growth behaviors, X and T. This data contains simulation that use different boundary conditions that either load the material vertically or horizontally, creating either horizontal or vertical fractures (respectively). We created four LLM embeddings describing these  scnearios, and we employed three of them for training (T-growth horizontal, T-growth vertical, X-growth horizontal) and the remaining one for testing (X-growth vertical). We trained a model to predict failure patterns and time to failure for the three training combinations, and we proved the ability of the model to generalize to the X-growth vertical case -- accurate zero-shot prediction.
In Table~\ref{tab:LLM-fp} we compare the accuracy on the prediction of the the failure pattern for the four cases and in Figure~\ref{fig:LLM-ttf} we show in a parity diagram the time to failure predictions for a test set made by 1000 samples. The model achieves over 96\% accuracy in predicting failure patterns across all cases in the test set, and the time-to-failure predictions demonstrate satisfactory performance for this pre-trained model. As shown in Figure~\ref{fig:diff_sizes}, the model is capable of learning the task regardless of image size (with edges ranging from 80 to 512 pixels) or the number of fractures (between 10 and 30) in the domain.
\section{Fine-tuning}
We employ a two-step fine-tuning process: first with phase-field simulations and then with finite discrete element method simulations. This fine-tuning step is critical, as phase-field and finite discrete element method are powerful physics-based models, offering a far more realistic representation of material failure than the rule-based models used in pre-training. While the rule-based data provides the model with foundational learning, fine-tuning on these high-fidelity simulations allows the model to capture more complex behaviors, ensuring it can generalize effectively to real-world scenarios.

\subsection{Phase-field simulations dataset}

In the first step of fine-tuning, we employ a phase-field method to generate fracture simulation data. The phase-field simulation method approximates fractures with a nonlinear diffusive phase-field \cite{Hirshikesh_etal_2019,Zhou_Zhuang_2020} driven by the accumulation of elastic strain energy. The phase-field introduces a degradation to an elastic medium as explained in Appendix~\ref{sec:phase-field}. The major advantage of this method is that it can represent complex fractures and fracture network naturally without explicitly specifying individual fractures. It can be formulated as explicit time-domain process or quasi-static process. In this work, we solve the fracturing propagation problem as a quasi-static process to improve the computational efficiency using the algorithm detailed in Appendix~\ref{sec:phase-field}.

We first generate random fracture configurations with orthogonal fractures randomly distributed in space, and set the initial fractures with a value of one as the initial condition of the phase-field. We then use this initial condition along with pulling boundary conditions where each model is stretched from the top and bottom simulate uniaxial tension. With the increase of displacement on the boundaries and within the model, the strain energy accumulates and the phase-field grows, mimicking the propagation of fractures. The growth of fractures stops until full failure of the model occurs, i.e., where there is connected fracture path between the boundaries of a model. 

Three materials, including steel, copper, and PBX-9501 (a polymer-bonded explosive), are simulated using this model. For each material, 50,000 simulations were conducted, each with a unique random initial fracture pattern. Of all the simulation data, 90\% were used to fine-tune the model, while 5,000 simulations were reserved for testing. Three new LLM embeddings have been created to characterize the materials and the phase-field model as source of the data.


\begin{figure}[htb!]
    \centering

    \begin{minipage}{0.6\textwidth}
        \centering
    \includegraphics[width=\textwidth]{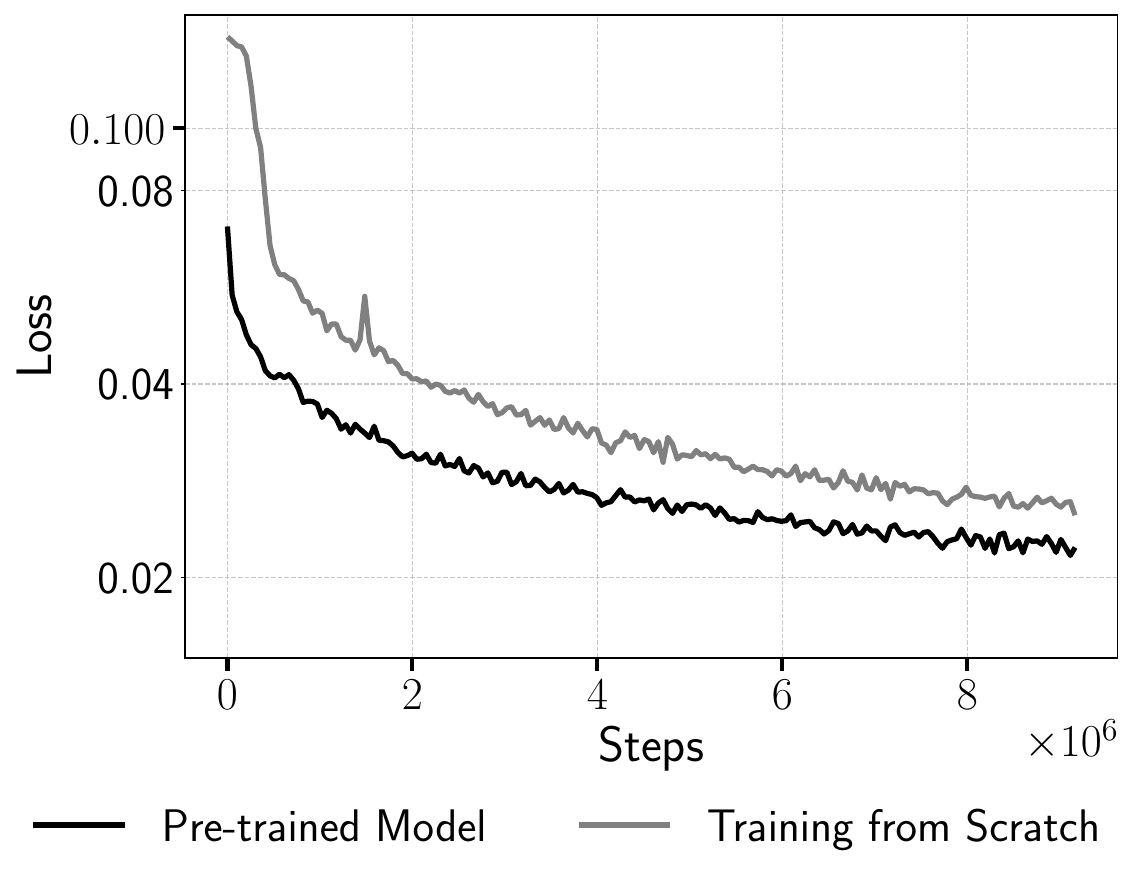}
    \caption{Comparison of the training loss of the pre-trained model vs training from scratch}
    \label{fig:loss_comparison}
    \end{minipage}
 \hspace{0.05\textwidth}
        \begin{minipage}{0.3\textwidth}
        \centering

\begin{tabular}{@{}ccc@{}}
\toprule
  Material      & Training set & Test set \\ \midrule
PBX    &    0.0189          &   0.0372       \\
Steel &      0.0271        &     0.0416     \\
Copper  &      0.0209        &    0.0290      \\ \bottomrule

\end{tabular}
 \captionof{table}{Mean cross-entropy loss for the training and the test set for three different materials.}

 \label{tab:materials_loss}
    \end{minipage}%
\end{figure}

\begin{figure}
    \centering
    \includegraphics[width=0.9\textwidth]{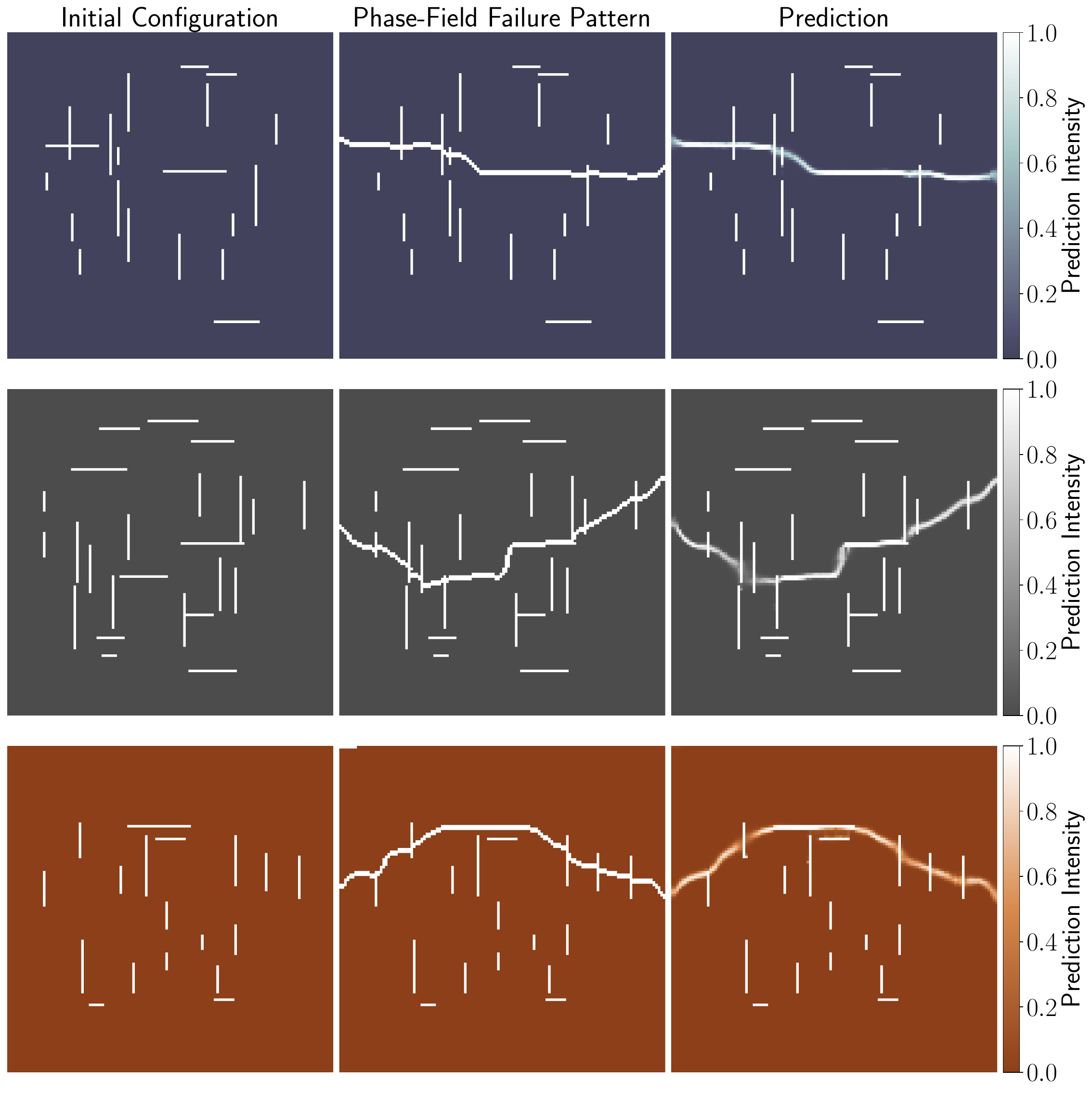}
    \caption{Comparison of the initial configuration, target, and prediction for PBX-9501, steel, and copper. The predictions spans between 0 to 1 as they are the raw output of a sigmoid function from the last layer of the network. }
    \label{fig:phase-field-pred}
\end{figure}

In Figure~\ref{fig:loss_comparison}, we present a comparison of the loss curves between a pre-trained model (trained on the rule-based data as explained in the previous section) and a model trained from scratch (default initialization in Pytorch). The pre-trained model consistently shows a lower loss, demonstrating an advantage over training from scratch. This advantage is particularly valuable when the available dataset is small, as pre-training helps mitigate overfitting. In order to evaluate our fine-tuned model, we employed cross-entropy as the evaluation metric since it allows us to directly assess the model's output without the need to set a threshold for the predictions. Table~\ref{tab:materials_loss} compares the cross-entropy values for both the training and test sets, and for the three different materials (PBX-9501, steel, and copper), showing that our model is not overfitting. Additionally, Figure~\ref{fig:phase-field-pred} provides examples of test set predictions for the three materials, the model predictions spans between 0 to 1 as they are the raw output from the last layer of the network, representing the model uncertainly in predicting the presence of a fracture. 


\subsection{Finite discrete elements simulations dataset}

In the second step of fine-tuning, we employ an advanced deformation and fracturing parallel solver, Hybrid Optimization Software Suite (HOSS) \cite{Knight_etal_2020}, to simulate how a material behaves under axial load. The simulation domain consists of a 0.25m by 0.25m plane strain sample that contains predefined horizontal and vertical fractures.  In the simulations, the sample was stretched from both the top and bottom at a constant speed of 1~m/s to simulate uniaxial tension. 

HOSS is particularly powerful because it not only can model the overall dynamic behavior of a material but also can accurately capture the fracture processes and interactions between different discrete parts of the material. In addition, the HOSS simulations are built on unstructured triangular meshes, which allow for easy incorporation of these preexisting fractures. The computational cost of HOSS simulation essentially depends on the size of the model, the mesh resolution being used, and the total simulation time. More details of the HOSS dataset generation and the corresponding simulation methods are included in Appendix ~\ref{sec:hoss}.
In order to work with unstructured simulation data, we implemented a positional encoder that takes as input the coordinates and the associated field values. This positional encoder can deal with both structured (e.g. images) and unstructured data as well as short- and long-range spatial relations.

\begin{figure}
    \centering
    \includegraphics[width=\textwidth]{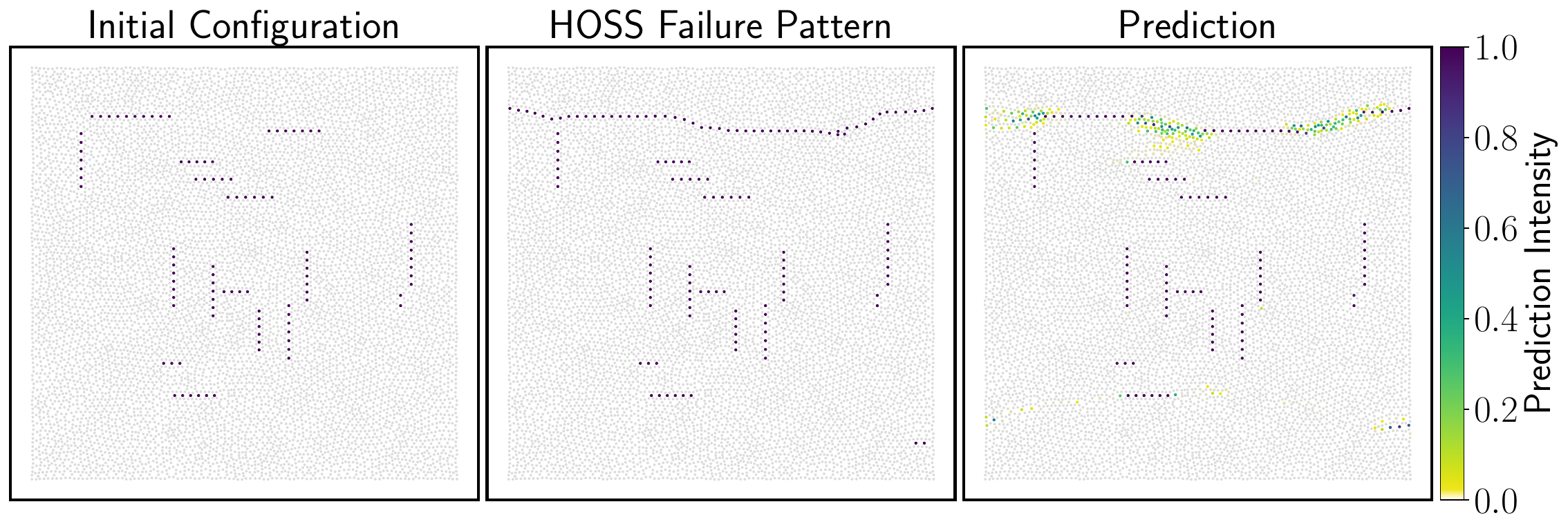}
    \caption{Comparison of the initial configuration, target, and prediction for a HOSS test example.}
    \label{fig:hoss_pred}
\end{figure}

We further fine-tuned the foundation model using the model fine-tuned on the phase-field simulations to initialize the parameters. This allows us to take advantage of the patterns learnt from the phase-field dataset training. In fact, we trained on 19000 computationally expensive simulations of PBX-9501, so the pre-training on the larger phase-field dataset is instrumental for the success for the HOSS fine-tuning of the model. We used 1000 HOSS simulations as our test set, one sample is shown in Figure \ref{fig:hoss_pred}, along with the corresponding model predictions. The overall cross-entropy for the training set is 0.01506, while the one for the test set is 0.0607. This indicates that our foundation model is performing well on the HOSS simulations, which have a much smaller dataset than the rule-based and phase-field simulators.


\section{Conclusions}

In this work, we introduced the first multimodal foundation model for predicting material failure, capable of handling a diverse range of input types and simulation conditions. Our proposed architecture demonstrates strong potential for generalization across various complex fracture datasets, moving beyond the limitations of previous models \cite{MOORE201846,srinivasan2018quantifying,hunter2019reduced,wang2021stressnet}, which were often restricted to specific material systems and boundary conditions. We found that increasing the number of parameters caused the loss to drop rapidly on our pre-training dataset -- much faster than that of language models. It remains an open question if this scaling will hold for more complex physics models. We hypothesize that as the physics becomes more complex and diverse (e.g., including nucleation, fluid flow, grain boundaries, plasticity, etc.), larger models will be needed to accurately reproduce these multiscale complex physics simulations. By leveraging large-scale data and a flexible, scalable architecture, our approach significantly improves the model's ability to handle both structured and unstructured grids, predict different fracture behaviors, and accommodate varying loading conditions.

While our results are preliminary, the use of large-scale models, embeddings, and physics-informed fine-tuning shows promise for broad applicability in material science and related fields. A truly general material fracture model would have broad applicability in engineering, geoscience, and beyond for design, optimization, and real-time predictions.



\bibliographystyle{abbrv}
\bibliography{refs}


\appendix

\section{Architecture details}
\label{sec:architecture}

The model's forward pass is composed of two primary components: an encoder and a decoder, which work together to process input data and generate desired outputs. The main hyperparameters for both the architecture and training process are outlined in Table~\ref{tab:hyperp}, while the parameters used to scale up the architecture are listed separately in Table~\ref{tab:scaleup}.

\begin{table}[htb!]

\vspace*{1em}
\centering
\begin{tabular}{@{}lc@{}}

\toprule
Hyperparameter & Value \\ \midrule
Space bands in positional encoder &    32   \\
Number of latents &     2048  \\
Number of cross attention heads (encoder) &     2  \\
Number of self attention heads (encoder) &     2  \\
Number of cross attention heads (decoder) &     1  \\
Number of layers & Tab.~\ref{tab:scaleup} \\  
Number of channels & Tab.~\ref{tab:scaleup} \\
Gradient accumulation&     64  \\
Gradient clipping&     0.5  \\
Weight decay&     0.1  \\
Learning rate (set-point) & 10$^{-4}$ \\

\bottomrule
\end{tabular}

\vspace*{1em}
\caption{Hyperparameters of architecture and training}
\label{tab:hyperp}
\end{table}

\subsection{Input Processing and Encoding}

The encoding process is carried-out as follows:

\begin{itemize}

     \item \textbf{Positional Encoding}: To help the model distinguish spatial locations, an $n$-dimensional positional encoding based on sine and cosine functions is concatenated with each pixel of the input data. For structured grids, this encoding aligns with the Cartesian coordinates. For unstructured grids (e.g., in finite element simulations), the positional encoding uses the coordinates of the element centers, allowing the model to capture the spatial context of the unstructured data.

     \item \textbf{Linear Layer}: The input data, which may come in different formats such as Cartesian or unstructured grids, is initially processed by a linear layer to fit the input's dimension with the rest of model's architecture. 

    \item \textbf{Latent Representation Initialization}: Learnable latent vectors are initialized and repeated across the batch to match the input size, forming the foundation of the model's internal representation.

    \item \textbf{Cross-Attention}: Cross-attention is applied between the input data and latent vectors, enabling the latent vectors to extract relevant information and refine the internal representation.

    \item \textbf{Self-Attention Layers}: The latent representation is further processed through self-attention layers, capturing complex relationships within the latent space. Residual connections stabilize training and enhance learning efficiency.
    
\end{itemize}

\subsection{LLM integration}

The decoder incorporates additional contextual information from a large language model (LLM) by embedding LLM-generated features into the latent space. This integration enhances the latent representation with nuanced, context-specific information, which is particularly valuable for tasks such as predicting material fractures or interpreting complex spatial behaviors.

\begin{figure}
    \centering
    \includegraphics[width=0.9\textwidth]{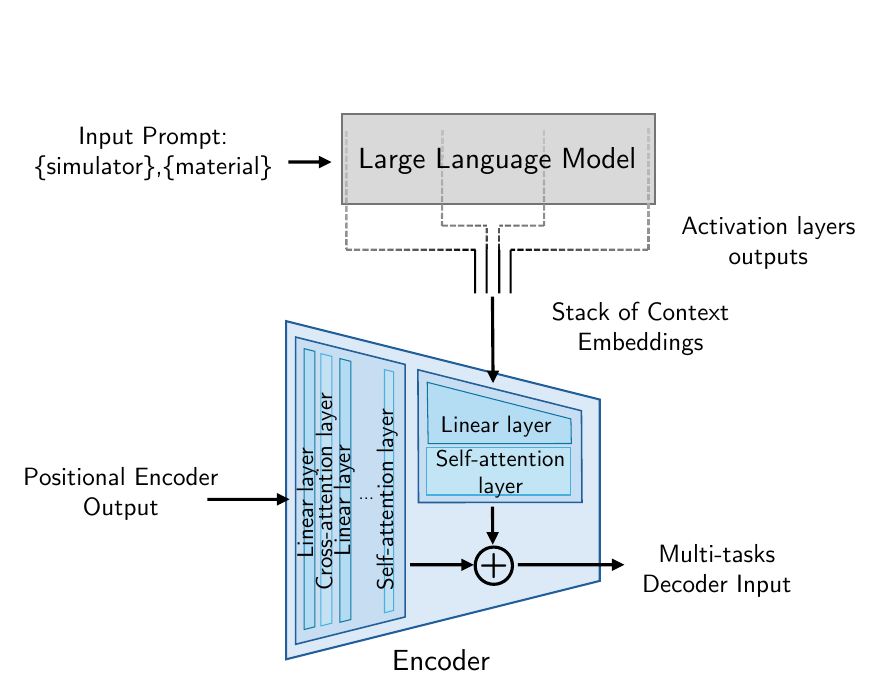}
    \caption{Schematic of the integration of the LLM embeddings in the encoder.}
    \label{fig:LLM-workflow}
\end{figure}

To integrate the LLM embeddings with the encoder output, we first applied a linear layer to reduce their dimensionality to number of channels of the encoder. After applying self-attention to this representation, we concatenated it with the encoder output, enabling more robust feature integration. A schematic illustrating this integration is shown in Figure~\ref{fig:LLM-workflow}.

\subsection{Decoding and Output Generation}

\begin{itemize}
    \item \textbf{Cross-Attention and Output Vectors}: The decoder applies cross-attention mechanisms to merge the enriched latent representation with specific output vectors. These vectors are divided into two categories:
    \begin{itemize}
        \item \textbf{Field Outputs}: These vectors generate spatial predictions, such as fields or images that represent the material's condition across a grid.
        \item \textbf{Scalar Outputs}: These vectors produce single-value predictions, such as time-to-failure estimates or other scalar quantities of interest.
    \end{itemize}
\end{itemize}

\subsection{Scaling up  the architecture}
\label{sec:scaleup}

Details of the scale-up of the models of Figure~\ref{fig:loss_accuracy}-~\ref{fig:scheduler}. We modified the number of channels in the encoder and in the decoder, and the number of self attention layers as in Table~\ref{tab:scaleup}. All the other parameters have been kept constant as detailed in the previous Appendix~\ref{sec:architecture}.

\begin{table}[htb!]

\vspace*{1em}

\begin{tabular}{@{}lcccc@{}}
\toprule
$\mathrm{N_{params}}$ & \multicolumn{1}{c}{\begin{tabular}[c]{@{}c@{}}Number of channels\\  encoder\end{tabular}} & \multicolumn{1}{c}{\begin{tabular}[c]{@{}c@{}}Number of channels\\  decoder\end{tabular}} & Number of self attention layers &  \\ \midrule
\textcolor[HTML]{000003}{\fontsize{13pt}{13pt}\textbullet} 103 k&    32   &   32     &   3    &  \\
\textcolor[HTML]{290b54}{\fontsize{13pt}{13pt}\textbullet} 441 k&     64  &   64     &   4    &  \\
\textcolor[HTML]{6a176e}{\fontsize{13pt}{13pt}\textbullet} 1.421 M&    128   &   128     &   3    &  \\
\textcolor[HTML]{a72d5f}{\fontsize{13pt}{13pt}\textbullet} 5.530 M&   256    &    256    &    3   &  \\
\textcolor[HTML]{dc5039}{\fontsize{13pt}{13pt}\textbullet} 21.808 M&    512   &    512    &   3    &  \\
\textcolor[HTML]{f98e08}{\fontsize{13pt}{13pt}\textbullet} 120.196 M&    1024   &   1024     &   5    &  \\
\textcolor[HTML]{345e8d}{\fontsize{13pt}{13pt}\textbullet} 815 M&    2048   &     2048   &    10   &  \\
\textcolor[HTML]{79d151}{\fontsize{13pt}{13pt}\textbullet} 3.258 B&   4096    &     4096   &   10    &  \\
\bottomrule
\end{tabular}

\vspace*{1em}
\caption{Scale-up of the models: hyperparameters tuned.}
\label{tab:scaleup}
\end{table}

\section{Creation of LLM embeddings}\label{sec:creation_emb}

We utilized a pre-trained large language model, Meta-Llama-3.1-8B-Instruct \cite{touvron2023llamaopenefficientfoundation}, to generate embeddings for various fracture scenarios involving different materials, fracture types, and directions of applied shear stress. This process allows us to capture the model's contextual understanding of these scenarios, which can aid the downstream  tasks.

We crafted a diverse set of prompts designed to describe different fracture scenarios. Each prompt incorporates a specific fracture type, material, and direction of applied shear stress. The prompts were formulated in multiple sentence structures to introduce variability in the input data and to ensure comprehensive coverage of potential scenarios.

To understand how the model processes these prompts, we extracted self-attention activations from selected layers within the model (the first layer, layers 7, 15, and the final layer). Self-attention activations provide insights into how the model attends to different components of the input sequence, revealing the internal mechanisms through which the model encodes the provided information.

For each unique combination of fracture type, material type, and shear stress direction, the corresponding prompt was fed into the model. As the model processed the prompt, the self-attention activations were captured and stored. These activations, representing the model's internal state at different layers, were saved as embeddings.

\section{Datasets}
\subsection{Rule-based algorithm}
\label{sec:wannabe}

This algorithm simulates the propagation of fractures in a 2D material grid over time and checks for material failure. It begins with initializing a material matrix where cells represent non-fractured (0) or fractured (1) points. Fractures propagate from specified initial points (called ``tips''), which move in a predefined direction each timestep (horizontal or vertical). The fracture tips have a freezing mechanism, which temporarily halt their movement when they encounter other fractures, depending on the simulation mode. In the T growth mode, the fractures stop their expansion when they encounter other fractures, while in the X growth mode, the fractures freeze temporarily then continue to propagate in the same direction. As fractures propagate, the material grid is updated to reflect fractured cells over multiple time steps. The fractures are then analyzed using graph theory, where the grid is represented as a graph, and neighboring fractured cells are connected by edges. The algorithm checks if there is a path across the material, either vertically or horizontally, which would indicate material failure. If such a path exists, the simulation records the time of failure. 

\begin{figure}
    \makebox[\textwidth][c]{
        \includegraphics[width=1.3\textwidth]{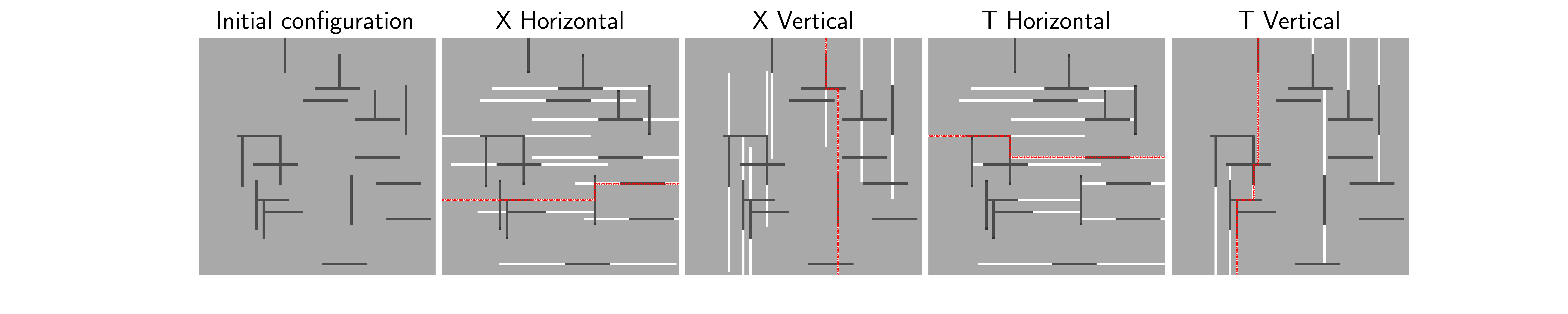}
    }
    \caption{Examples of outcomes of the rule-based algorithm. Initial configuration of the fractures, Failure pattern for the four different simulation modes: X horizontal, X vertical, T horizontal, T vertical. In red the path connecting the boundaries of the material (right-left for horizontal growth, top-bottom for vertical growth) causing the material failure.}
    \label{fig:wannabe}
\end{figure}

Failure patterns for X or T, horizontal or vertical, are shown in Figure~\ref{fig:wannabe}. 

\subsection{Numerical simulation of fracture propagation using the phase-field  method}\label{sec:phase-field}

In the phase-field fracture propagation theory, the elastic constitutive relation reads \citep{Hirshikesh_etal_2019,Zhou_Zhuang_2020}:
\begin{equation}
\sigma (\bfu)  = \left[(1-\xi) (1 - \psi)^2 + \xi\right] \bfc: \left(\nabla \bfu + \nabla \bfu^{\mathrm{T}}\right), 
\label{eq:stress}
\end{equation}
where $\bfc = \bfc(\bfx) = c_{ijkl}(\bfx)$ is the fourth-order elasticity tensor of the matrix with up to 21 independent components; $\xi$ is a small positive number to avoid instability at the fracture $\psi$, The Neumann boundary condition for $\sigma$ is 
$\sigma \cdot \bfn = \bft_N$, where $\bfn$ is the normal vector of the Neumann boundary $\partial \Omega_{N}$ of the computational domain $\Omega$, and $\bft$ is an external force. 

The evolution of the fracture phase-field $\phi$ is driven by the so-called positive elastic strain energy. Based on the principle of variation, the governing equation for fracture evolution can be written as \cite{Hirshikesh_etal_2019,Zhou_Zhuang_2020}:
\begin{equation}
	G_c \left(w_0^{-1} \psi - w_0 \nabla^2 \psi \right) = 2(1- \psi) H^+(\varepsilon, t),
	\label{eq:pf}
\end{equation}
where $2w_0$ is the characteristic width of a fracture and $G_c$ is the critical energy release rate; $H^+(\varepsilon, t)$ is a history energy field defined as the maximum positive elastic energy at $\bfx$ up to $t$: 
\begin{equation}
	H^+(\varepsilon, t)  = \max_{\tau \in [0, t]} \Phi^+(\varepsilon, \tau). 
\end{equation}
While there are a number of choices to define the positive elastic strain energy function $\Phi^+$, here we choose one based on the elastic tensor $\bfc$ in a general form for computing the following elastic energy: 
\begin{equation}
	\Phi^+ (\varepsilon, t) = \frac{1}{2} (\varepsilon^+)^{\mathrm{T}} : \mathbf{c} :   \varepsilon^+ ,
	\label{eq:history}
\end{equation}
where the tensile-positive elastic strain tensor reads
\begin{equation}
	\varepsilon^+ = \left[\varepsilon_{xx}^+, \varepsilon_{yy}^+, \varepsilon_{zz}^+, \varepsilon_{yz},   \varepsilon_{xz},  \varepsilon_{xy}\right]^{\mathrm{T}}, 
\end{equation}
with the tensile-positive strain tensor defined as
\begin{equation}
	\varepsilon_{ij}^+ = \frac{1}{2} \left(\varepsilon_{ij} + |\varepsilon_{ij}|\right),
\end{equation}
where $\varepsilon_{ij} = \frac{1}{2}(\partial_i u_j + \partial_j u_i)$ is the strain tensor. The above definition allows us to define the history energy function for both isotropic and anisotropic materials in a unified formulation. 

We solve the above fracture evolution system using our in-house solver. In specific, we use spectral elements \cite{Komatitsch_Tromp_1999} to discretize the elastic constitutive equation and the phase-field evolution equation, where the unknowns are vector particle displacement field $\bfu$ and the scalar phase-field $\phi$, respectively. We use an iterative Krylov subspace solver along with a symmetric successive over-relaxation (SOR) preconditioner to solve the associated sparse linear systems provided by the numerical library PETSc \cite{Balay_etal_1998}. 

To generate fracture simulation data with the above phase-field method, we first generate a total of one million 2D random fracture models, where we place a random number of randomly distributed fractures inside a square domain with a dimension of 0.25~m by 0.25~m. The orientations of the fractures are either 0 or $\pi/2$. Therefore, the resulting fractures are orthogonal in terms of orientation, but not all of them are overlapping. In this application, we assume varying Dirichlet boundary conditions, $\bfu_z^+ = t$~m and $\bfu_z^- = -t$~m, where $t$ is the time, as the pulling forces along the vertical directions of the domain. We discretize the model domain with a total of $127\times 127$ regular first-order spectral elements, resulting in a total of $128\times 128$ degrees of freedom for the phase-field equation~\ref{eq:pf} and a total of $2\times 128\times 128$ unknowns for the elastic constitutive equation~\ref{eq:stress} in the 2D scenario. 

\begin{figure}[htb!]
\centering
\subfloat[]{\includegraphics[height=0.175\textheight]{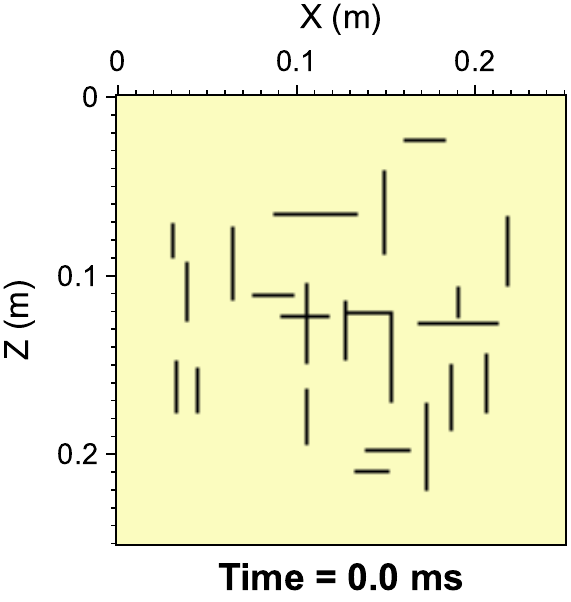}}
\subfloat[]{\includegraphics[height=0.175\textheight]{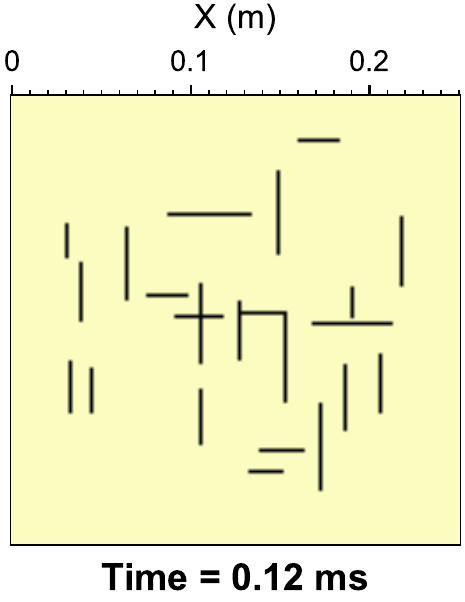}}
\subfloat[]{\includegraphics[height=0.175\textheight]{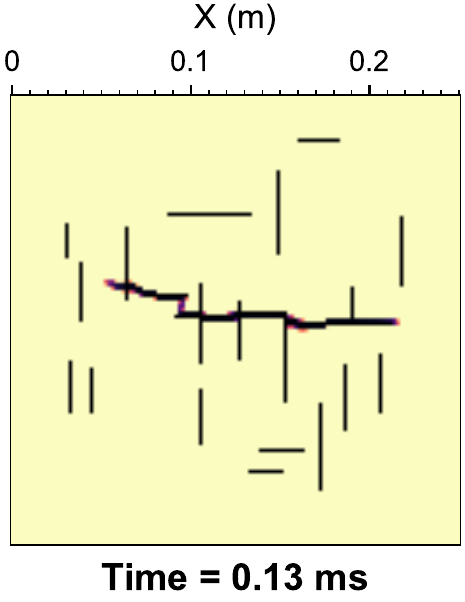}}
\subfloat[]{\includegraphics[height=0.175\textheight]{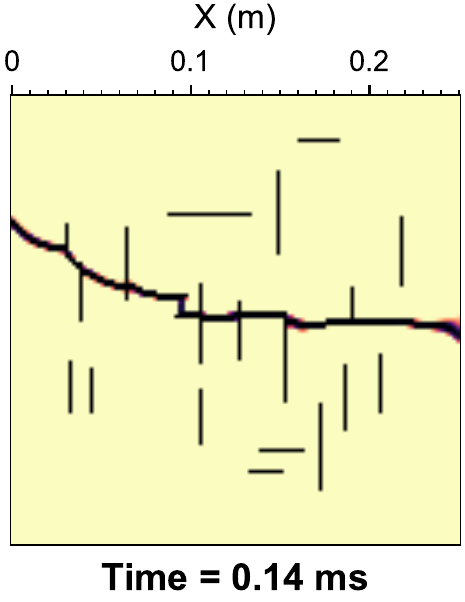}} \\
\vspace*{-1em}
\subfloat[]{\includegraphics[height=0.175\textheight]{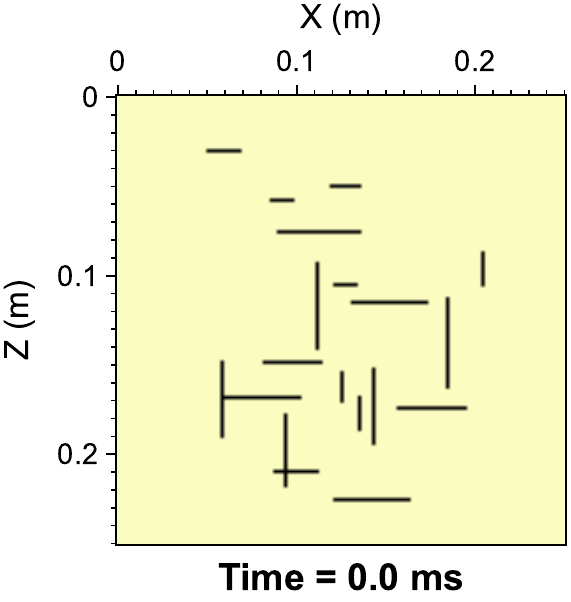}}
\subfloat[]{\includegraphics[height=0.175\textheight]{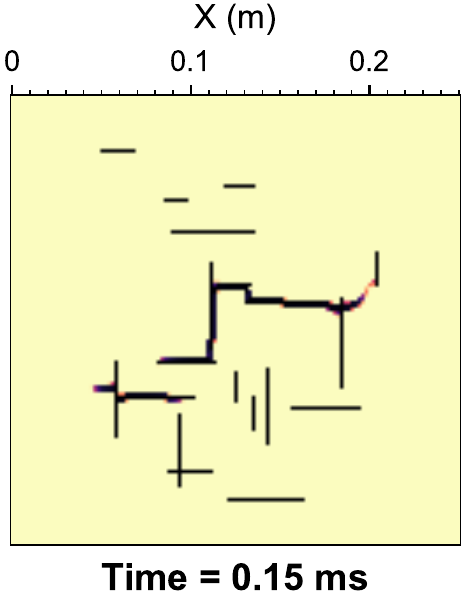}}
\subfloat[]{\includegraphics[height=0.175\textheight]{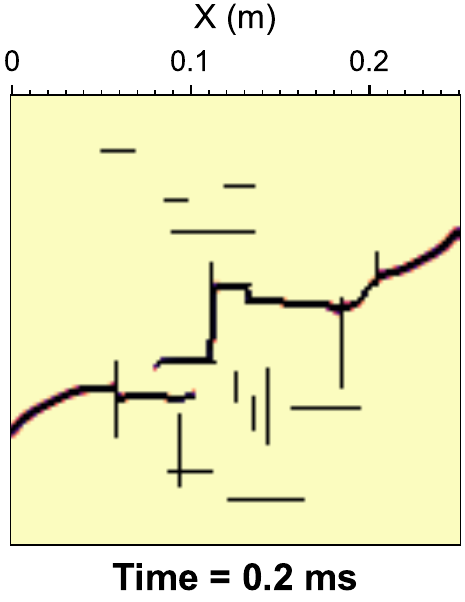}}
\subfloat[]{\includegraphics[height=0.175\textheight]{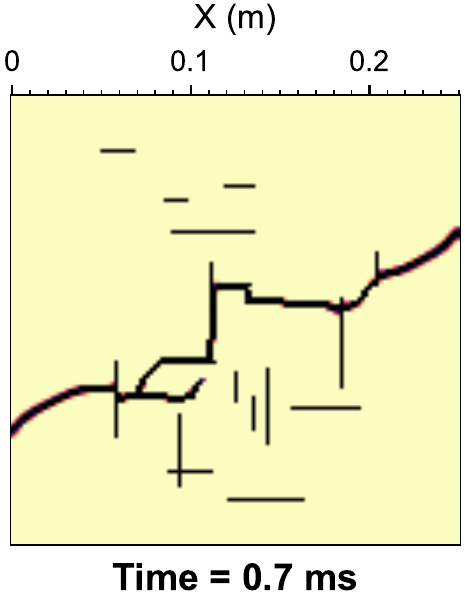}} 
\caption{Two examples of random fracture models (a, e) and fracture simulation results (b-d, f-h) at different time. Note that the failure time (where full breakage of material occurs) of these two fracture models are different.}
\label{fig:pf_frac}
\end{figure}

Figure~\ref{fig:pf_frac} displays two examples of the generated random fracture models and the snapshots of fracture evolution under the prescribed boundary conditions for a material, PBX-9501. The material has a Young's modulus of 10~GPa, a Poisson's ratio of 0.36, and a density of 1,820~kg/m$^3$. We adopt a critical energy release rate of 641~N/m. The results indicate the strong nonlinearity and complexity of fracture evolution of this material. For both materials, the fracturing occurs after a certain amount of vertical displacement of boundaries, and quickly propagate to model boundaries. For the first model, it reaches full breakage (i.e., material failure) at 0.14~ms. However, for the second random fracture model, the failure does not occur until 0.7~ms, although a partial failure is observed at 0.2~ms. The results indicate that even for the same material and with the same boundary condition, the pattern of preexisting fractures may result in notable differences in the final failure pattern and failure time. From a physics simulation point of view, there is not a simple rule to determine the failure time and final pattern unless a simulation is conducted. This is where our work aims to achieving through the aforementioned foundation model trained by a large amount of such fracture simulation data. 

In this work, we performed fracturing simulations for a total of three materials: PBX-9501, steel, and copper. The elasticity parameters of the three materials are detailed in Table \ref{tab:materials}. 

\begin{table}[htb!]
\begin{centering}

\begin{tabular}{ccccccc}
\toprule 
Material & $C_{11}$ (GPa) & $C_{13}$ (GPa) & $C_{33}$ (GPa) & $C_{55}$ (GPa) & $\rho$ (kg/m$^3$) & $G_{c}$ (J/m$^2$)\tabularnewline
\midrule 
PBX-9501  & 15.9 & 8.5 & 15.9 & 3.7 & 1,820 & 641 \tabularnewline
Steel  & 282.7 &121.2 & 282.7 & 80.8  & 7,850 & 105 \tabularnewline
Copper  & 168.4 & 121.4 & 168.4 & 75.4 & 8,960 & 20 \tabularnewline
\bottomrule
\end{tabular}
\vspace{1em}
\caption{Properties of the materials used in the simulation. $C_{11}$, $C_{13}$, $C_{33}$, and $C_{55}$ are elasticity matrix parameters of the material in the Voigt notation; $\rho$ is the mass density, and $G_c$ is the Griffith critical energy release rate of fracture. }
\label{tab:materials}
\par 
\end{centering}
\end{table}

Currently, our in-house phase-field simulator does not consider plasticity (or ductility) of the materials, but only consider elasticity (brittleness) of the materials. However, since plasticity is an important mechanical property of many materials, particularly of metals, we are currently working on including both elasticity and plasticity into the phase-field solver with models such as \cite{Seles_etal_2021}. Nevertheless, we hypothesize that the efficacy and accuracy of our aforementioned foundation model will remain intact. 

\subsection{Numerical simulation of fracture propagation using the combined finite discrete-element method}\label{sec:hoss}

The Hybrid Optimization Simulation Suite (HOSS) \cite{Knight_etal_2020} is an advanced computational framework designed for simulating complex material behaviors, including fracturing and fragmentation. Researchers use HOSS due to its robustness, advanced algorithms, massive parallelization features, and versatile capabilities. HOSS results have been validated by numerous independent experiments \cite{Knight_etal_2020}. The non-exhaustive list of HOSS applications include the following: experimental rock mechanics (e.g. triaxial uniaxial), wellbore and drilling stability, building integrity, seismic studies, high velocity impacts, thermal-hydraulic-mechanic behavior, high explosive performance, tissue and cell experiments, and weapons penetration.

HOSS employs the combined finite-discrete element method \cite{munjiza_combined_2004} which merges finite element based analysis of continua with discrete element based transient dynamics, contact detection, and contact interaction solutions. The solid domain is divided into finite elements, assuming finite rotations and displacements, and analyzed using a multiplicative decomposition for finite strains \cite{munjiza_large_2015}. Here, composite triangular elements use selective stress integration to prevent artificial stiffness or locking. They also employ a unified hypo/hyper-elastic approach that accepts user-defined isotropic or anisotropic material models\cite{lei2016generalized,lei2016non}. When failure, fracture, or fragmentation occurs, a single finite element mesh breaks into multiple interacting domains. The same finite element approach is used to handle the contact between these discrete elements\cite{munjiza_compmechdis_2011}. In-house state-of-the-art discretized contact solutions handle contact detection and contact interaction\cite{Knight_etal_2020,lei2014framework}

\begin{figure}
\centering
\includegraphics[height=0.375\textheight]{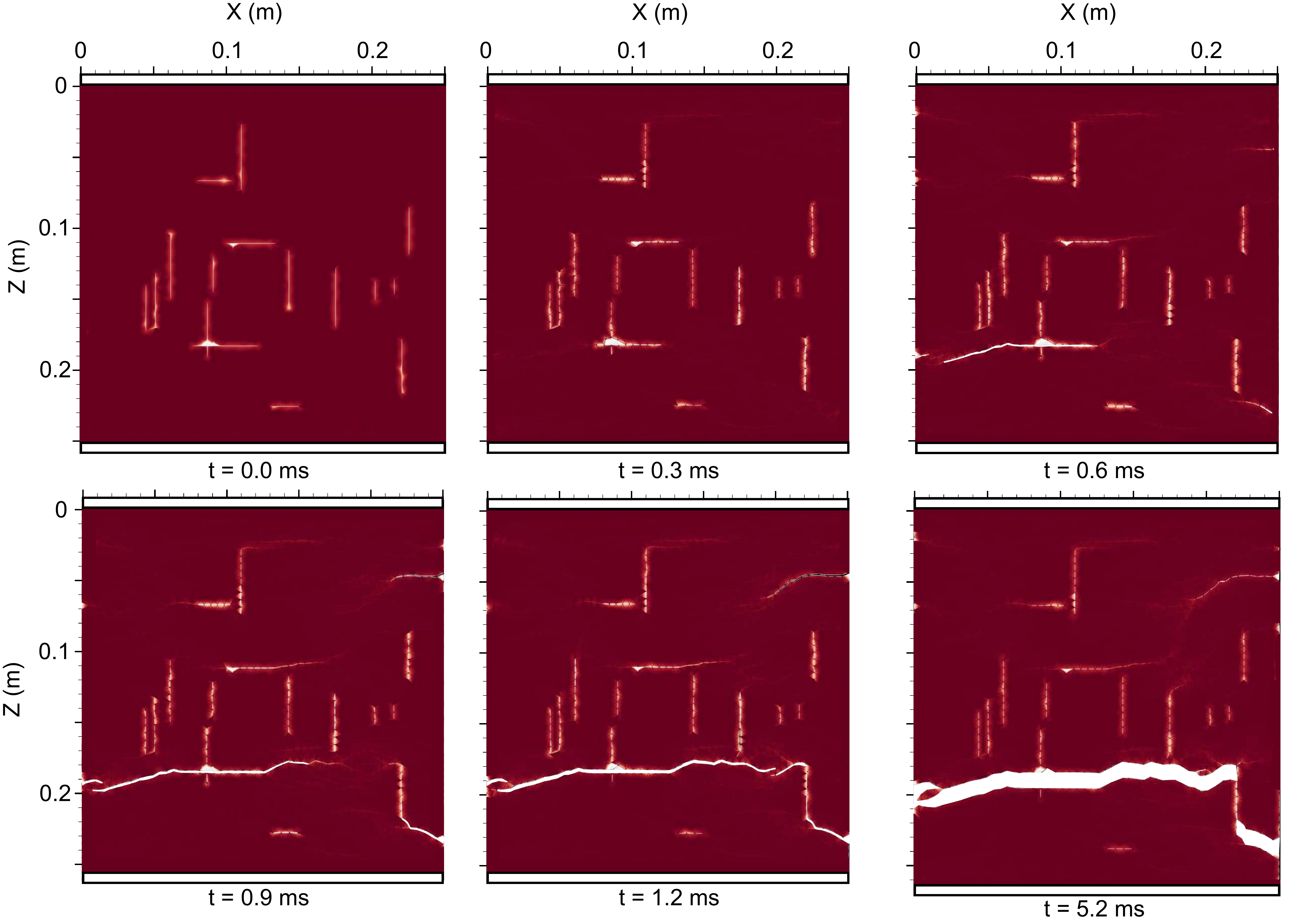}
\caption{Single HOSS fracture example at different time steps. Note, time to failure could be earlier in the simulation but requires post-processing from the final fracture pattern.}
\label{fig:hoss_frac}
\end{figure}

To generate fracture simulation data we first incorporate the same initial seeded fractures generated from \ref{sec:phase-field}. The randomly distributed fractures are placed in our 2D square domain that is 0.25 m by 0.25 m that is also confined on the top and bottom by 5 mm plates (Figure~\ref{fig:hoss_frac}). We assume free boundary conditions for the sides of the model and use a linear ramping (0 to $1\cdot10^{-6}$ s) nodal velocity boundary conditions for the plates that are $\pm$1 m/s (pulling). We use a fixed time increment of $1\cdot10^{-8}$ s. As we can not anticipate the exact time to failure of the material during the simulation we allow each simulations to run for a specified wall-clock time of 1 hour and 30 minutes using 8 MPI domains and CPU cores, which is the typical duration needed to achieve full failure. Current simulations use the same isotropic material properties and cohesion terms of a typical brittle rock. With increasing ML performance new materials and varied conditions will be implemented and trained on.

Since the simulations produce displacement between fractures the initialized domain increases in size vertically (Figure~\ref{fig:hoss_frac}). These results produce an unstructured grid of information that may also be long past the first time to failure step. To resolve these difficulties results are post-processed using the python pyvista package\cite{pyvista} which retains the edges of the elements that fractured in the earliest failure time while simultaneously retaining the same grid of the initialized domain.
Similar to the phase-field results (\ref{sec:phase-field}) there are a host of complexities that make the results suitable for foundation model training. Strong nonlinearity and complex fracture patterns emerge due to small perturbations in preexisting fracture patterns, which cannot be determined ad hoc.

\end{document}